\documentclass[aps,prl,groupedaddress,showpacs,preprintnumbers]
              {revtex4}%
\usepackage{graphicx}
\usepackage{amsmath}
\usepackage{bm}
\usepackage{relsize}
\RequirePackage{xspace}
\newcommand{\babar}{\mbox{\slshape B\kern-0.1em{\smaller A}\kern-0.1em %
B\kern-0.1em{\smaller A\kern-0.2em R}}\xspace}

\begin{document}

\preprint{\begin{tabular}{l}
          ANL-HEP-PR-02-110\\
          FERMILAB-PUB-02/332-T \end{tabular}}
\title{\mbox{}\\[10pt]
$\bm{e^+ e^-}$ Annihilation into $\bm{J/\psi + J/\psi}$}


\author{Geoffrey T.~Bodwin}
\affiliation{
High Energy Physics Division, 
Argonne National Laboratory, 
9700 South Cass Avenue, Argonne, Illinois 60439}

\author{Eric Braaten}
\affiliation{
Physics Department, Ohio State University, Columbus, Ohio 43210}
\affiliation{
Fermi National Accelerator Laboratory, P.~O.~Box 500, Batavia, Illinois 60510}

\author{Jungil Lee}
\affiliation{
Department of Physics, Korea University,
Seoul 136-701, Korea}


\date{\today}
\begin{abstract}
Recent measurements by the Belle Collaboration of the
exclusive production of two charmonia in $e^+ e^-$ annihilation
differ substantially from theoretical predictions. We suggest that a
significant part of the discrepancy can be explained by the process $e^+
e^- \to J/\psi + J/\psi$. Because the $J/\psi + J/\psi$ production
process can proceed through fragmentation of two virtual photons into two
$c\bar c$ pairs, its cross section may be larger than that for $J/\psi +
\eta_c$ by about a factor 1.8, in spite of a suppression factor
$\alpha^2/\alpha_s^2$ that is associated with the QED and QCD coupling
constants.
\end{abstract}

\pacs{13.66.Bc, 12.38.Bx, 14.40.Gx}

\maketitle


The charmonium states have long been a key testing ground for our
understanding of quantum chromodynamics (QCD), both because of their
clear experimental signatures and because of the theoretical
simplifications that arise from their nonrelativistic nature. The huge
data sample of continuum $e^+ e^-$-annihilation events now being
produced at the SLAC and KEK $B$ factories makes it possible to
investigate charmonium-production processes that have very small cross
sections. The exclusive production of double-charmonium states is a
particularly interesting example because the theoretical predictions do
not involve any unknown nonperturbative parameters. However, the first
such measurements by the Belle Collaboration \cite{Abe:2002rb} are in
substantial disagreement with existing predictions. In this paper, we
argue that a significant part of the discrepancy between experiment and
theory may be attributable to the process $e^+ e^-$ annihilation into
$J/\psi$ +$J/\psi$, where $J/\psi$ is the lowest spin-triplet charmonium
state. This process had been overlooked because it is suppressed by
$\alpha^2/\alpha_s^2$, where $\alpha$ is the quantum-electrodynamic
(QED) coupling, and $\alpha_s$ is the QCD coupling. However, as we shall
show, this process is enhanced by a kinematic factor that is associated
with the fragmentation of photons into $c \bar c$ pairs.

The Belle Collaboration has observed $e^+ e^-$ annihilation into two
charmonium states at a center-of-mass energy $\sqrt{s} = 10.6$ GeV
by studying the recoil-momentum spectrum of the $J/\psi$
\cite{Abe:2002rb}.  The collaboration measured the production cross
section for $J/\psi + \eta_c$ and also found evidence for $J/\psi +
\chi_{c0}$ and $J/\psi + \eta_c(2S)$ final states. Recent calculations
of the production cross section for $J/\psi + \eta_c$ have given results
that are about an order of magnitude smaller than the Belle measurement
\cite{Braaten:2002fi,Liu:2002wq}. This presents a challenge to our
current understanding of charmonium production based on NRQCD. 

\begin{figure}
\includegraphics[height=8cm,angle=-90]{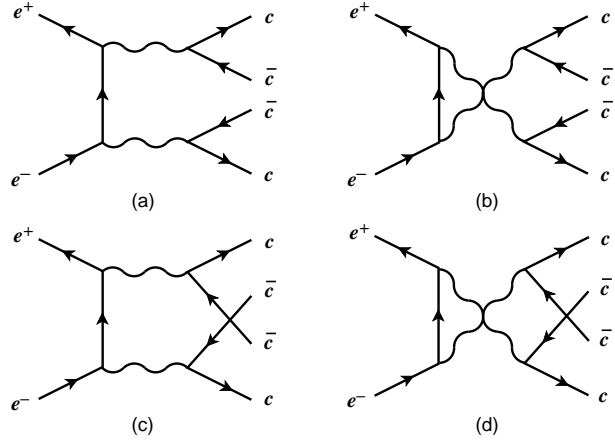}
\caption{\label{fig1}%
QED diagrams for $e^+ e^- \to \gamma^* \gamma^* \to c \bar c + c
\bar c$. The upper and lower $c\bar{c}$ pairs evolve into $H_1$ and
$H_2$, respectively.}
\end{figure}

States consisting of two charmonia
with opposite charge conjugation, such as
$J/\psi + \eta_c$, can be produced at order $\alpha^2 \alpha_s^2$
through processes $e^+ e^- \to c \bar c + c \bar c$ that involve only a
single virtual photon. 
(See Refs.~\cite{Braaten:2002fi,Liu:2002wq} for examples.)
States consisting of two charmonia
with the same charge conjugation, such as
$J/\psi + J/\psi$, can be produced at order $\alpha^4$ through the
processes shown in the diagrams in Fig.~\ref{fig1}, which involve two
virtual photons. 
We find that, in spite of its being suppressed by a
factor of $\alpha^2/\alpha_s^2$, the $J/\psi+J/\psi$ production cross
section may be larger than that for $J/\psi + \eta_c$. We suggest that some
of the events in Belle's $J/\psi+ \eta_c$ signal may actually be
$J/\psi+J/\psi$ events. By taking this effect into account, one would
decrease the discrepancy between the Belle measurements and the
predictions of NRQCD.

The effective field theory NRQCD can be used to
write a quarkonium production cross section as a sum of products of
short-distance coefficients and NRQCD matrix elements \cite{BBL}. The
short-distance coefficients can be calculated in QCD perturbation
theory, but the matrix elements are nonperturbative in nature. In the
nonrelativistic limit, there is a single independent matrix element for
each spin multiplet of charmonium states. The matrix elements can be
determined phenomenologically from electromagnetic annihilation decay
rates. Thus, the cross sections for double-charmonium production can be
predicted up to relativistic corrections without any unknown
nonperturbative factors.

It is convenient to express the cross sections for double-charmonium states 
in terms of the ratio $R$, which is defined by
\begin{eqnarray}
R[H_1 + H_2] = 
\frac{\sigma[e^+ e^- \to H_1 + H_2]}{ \sigma[e^+ e^- \to \mu^+ \mu^-]},
\label{R-def}
\end{eqnarray}
where $\sigma[e^+ e^- \to \mu^+ \mu^-]=\pi\alpha^2/(3E_{\rm beam}^2)$,
and $E_{\rm beam}=\sqrt{s}/2$. We define
the angular variable $x = \cos \theta$, where $\theta$ is the angle
between the $J/\psi$ and the beam in the center-of-mass frame. We also
introduce a dimensionless kinematic variable $r= 2 m_c/E_{\rm beam}$, 
with $m_c$ the charm-quark mass. In
the nonrelativistic limit, cross sections for the $1S$ states $J/\psi$
and $\eta_c$ have a common NRQCD matrix element $\langle O_1
\rangle_{1S}$ that is defined in Ref.~\cite{Braaten:2002fi}.

The cross section for $J/\psi + J/\psi$ receives contributions from
the diagrams in Fig.~\ref{fig1}. The angular distribution is
\begin{eqnarray}
\frac{dR}{dx}[J/\psi + J/\psi] &=& 
\frac{16\pi^2  \alpha^2 }{243m_c^6}  \,
\langle O_1 \rangle_{1S}^2
\nonumber\\
&& \hspace{-2cm} \times
\frac{(1-r^2)^{1/2}F(x)}
     {[4(1-r^2)(1-x^2)+r^4]^2},
\label{dRdx:psi+psi}
\end{eqnarray}
where 
\begin{eqnarray}
F(x)&=&r^4 x^2 (1-x^2) [ 2 + 3 r^4 -r^6 +4 x^2 (1-r^4) ]^2
+2(1+x^2) (1-x^2) 
\nonumber\\ &&\times
[ 6 - 7 r^2 +4 r^4-r^6 +4 x^2 r^2 (1-r^2) ]^2
+2r^4 x^2 (1-x^2) 
\nonumber\\ &&\times
[3 - 4 r^2 +4 r^4-r^6 +4 x^2 r^2 (1-r^2)  ]^2
+2r^2 [ 
  (2-r^2)^2 (3- 2 r^2 + r^4)^2
\nonumber\\ &&
-x^2 ( 72 - 228 r^2 
+ 359 r^4 - 292 r^6 + 130 r^8 -  32 r^{10} + 3 r^{12} )
\nonumber\\ &&
+4 x^4 ( 9 - 60 r^2 +122 r^4 
-112 r^6 + 58 r^8 - 14 r^{10} +    r^{12} )
\nonumber\\ &&
+16 x^6 r^2 (1-r^2) ( 6- 11 r^2+ 11 r^4
- 2 r^6)
+64 x^8 r^4 (1-r^2)^2
].
\end{eqnarray}
The ratio $R$ is obtained by integrating $x$ only from $0$ to $1$, in
order to avoid double-counting of identical final-state particles. The
angular distribution $dR/dx$ for $J/\psi + \psi(2S)$ is given by an
expression identical to that in Eq.~(\ref{dRdx:psi+psi}), except that
one of the factors of $\langle O_1 \rangle_{1S}$ is replaced by $\langle
O_1 \rangle_{2S}$, and the range of $x$ is from $-1$ to $1$.

The cross section for $\eta_c + \eta_c$ receives contributions only from
the two diagrams in Fig.~\ref{fig1}(c) and \ref{fig1}(d).
The angular distribution is
\begin{eqnarray}
\frac{dR}{dx}[\eta_c + \eta_c] &=&
\frac{ 16\pi^2 \alpha^2 }{243m_c^6} \,
\langle O_1 \rangle_{1S}^2 
\nonumber
\\
&& 
\times \, r^4 (1-r^2)^{5/2}x^2(1-x^2).
\label{dRdx:eta+eta}
\end{eqnarray}
The ratio $R$ is obtained by integrating $x$ from $0$ to $1$.
The angular distribution $dR/dx$ for $\eta_c + \eta_c(2S)$ is 
identical to that in Eq.~(\ref{dRdx:eta+eta}), except
that one of the factors of $\langle O_1 \rangle_{1S}$ is replaced by 
$\langle O_1 \rangle_{2S}$, and the range of $x$ is $-1$ to $1$.

The production of $J/\psi+\eta_c$ proceeds through $e^+ e^-$
annihilation into a single virtual photon, which creates a $c\bar c$
pair. The second $c\bar c$ pair is then created either by a virtual gluon
or by a virtual photon radiated from the first $c\bar c$ pair. The cross
section was recently calculated by Braaten and Lee
\cite{Braaten:2002fi}. The differential ratio is
\begin{eqnarray}
\frac{dR}{dx}[J/\psi + \eta_c]
&=&
\frac{4\pi^2 }{2187m_c^6} \,\langle O_1 \rangle_{1S}^2
[3 \alpha_s r^2 + \alpha(3+r^2)]^2
 r^2 (1-r^2)^{3/2} (1+x^2).
\end{eqnarray}
The ratio $R$ is obtained by integrating over $x$ from $-1$ to $1$. The
$\alpha_s^2$ term was also calculated recently by Liu, He, and Chao
\cite{Liu:2002wq}. It had been calculated previously, but the analytic
expression was not given \cite{Brodsky:1985cr}. The QED terms 
increase the cross section by about 29\% for 
$E_{\rm beam} = 5.3$ GeV.

When $E_{\rm beam}$ is much greater than $m_c$, the relative sizes of the
double-charmonium cross sections are governed not only by the powers of
the coupling constants, but also by the number of kinematic suppression
factors $r$. The power-counting rules of perturbative QCD \cite{pqcd}
require $dR/dx$ to behave as $r^4$ in the limit $r \to 0$, with a further
suppression factor of $r^{2|\lambda_1 + \lambda_2|}$ if the sum of the
charmonium helicities, $\lambda_1$ and $\lambda_2$, does not vanish. In
the limit $r \to 0$, $dR/dx$ for $\eta_c + \eta_c$ scales as $\alpha^2
r^4$, while $dR/dx$ for $J/\psi + \eta_c$ scales as $\alpha_s^2 r^6$,
with the extra factor of $r^2$ arising from helicity suppression. As $r
\to 0$ with $x$ fixed, $dR/dx$ for $J/\psi + J/\psi$ approaches a
constant. The power-counting rules of perturbative QCD are evaded
because this process has a contribution, corresponding to the diagrams
of Figs.~\ref{fig1}(a) and \ref{fig1}(b), in which two virtual photons,
with virtuality of order $m_c$, fragment into two $c \bar c$ pairs. This
contribution is enhanced because the virtual-photon propagators are of
order $1/m_c^2$ instead of order $1/E_{\rm beam}^2$. In the amplitude,
there are also two numerator factors of $m_c$ instead of $E_{\rm beam}$,
which arise from the $c\bar c$ electromagnetic currents. Hence, the net
enhancement of the squared amplitude is $(E_{\rm beam}/m_c)^4$. The
integrated ratio is further enhanced by a factor $\ln(1/r)$ because the
potential logarithmic divergence in the integral of
Eq.~(\ref{dRdx:psi+psi}) as $x\to1$ is cut off at $1-x^2 \sim r^4$.

Chang, Qiao, and Wang have pointed out that the cross section for $e^+
e^- \to J/\psi + \gamma$ is enhanced if $E_{\rm beam}$ is much greater
than $m_c$ \cite{Chang:1997dw}. The reason for the enhancement is that
this process also involves a contribution in which a virtual photon
fragments into a $c \bar c$ pair.

We use the results given above to predict the cross sections for
producing two $S$-wave charmonium states at the $B$ factories, with
$E_{\rm beam} = 5.3$ GeV. The ratios $R$ depend on the coupling
constants $\alpha$ and $\alpha_s$, the charm-quark mass $m_c$, and the
NRQCD matrix elements. We take the QCD coupling constant to be $\alpha_s
= 0.21$, which corresponds to the renormalization scale 5.3 GeV. For
$m_c$, we use the next-to-leading order pole mass, which we take to be
$m_c = 1.4 \pm 0.2$~GeV. The NRQCD matrix elements can be determined
from the electromagnetic annihilation decay rates of $J/\psi$ and
$\psi(2S)$. We use the values from the analysis of
Ref.~\cite{Braaten:2002fi}. For $m_c = 1.4$ GeV, they are $\langle O_1
\rangle_{1S} = 0.335 \pm 0.024$ GeV$^3$ and $\langle O_1 \rangle_{2S} =
0.139 \pm 0.010$ GeV$^3$, where the error bars are those associated with
the experimental uncertainties only. For other values of $m_c$, the
NRQCD matrix elements should be multiplied by $(m_c/1.4 \, {\rm GeV})^2$.

\begin{table}[hb]
\caption{\label{tab:sigmaSS}
Cross sections in fb for $e^+ e^-$ annihilation into two $S$-wave
charmonium states $H_1+H_2$ at $E_{\rm beam} = 5.3$~GeV. The errors are
only those from variations in the pole mass $m_c = 1.4 \pm 0.2$ GeV.
There are additional large errors associated with perturbative-QCD  and
relativistic corrections, as described in the text.}
\begin{ruledtabular}
\begin{tabular}{ll}
$H_1+H_2$&$\sigma$~(fb)\\
\hline
$J/\psi  +J/\psi$      & 6.65 $\pm$ 3.02
\\
$J/\psi  +\psi(2S)$    & 5.52 $\pm$ 2.50
\\
$\psi(2S)+\psi(2S)$    & 1.15 $\pm$ 0.52
\\
\hline
$J/\psi  +\eta_c$      & 3.78 $\pm$ 1.26
\\
$J/\psi  +\eta_c(2S)$  & 1.57 $\pm$ 0.52
\\
$\psi(2S)+\eta_c$      & 1.57 $\pm$ 0.52
\\
$\psi(2S)+\eta_c(2S)$  & 0.65 $\pm$ 0.22
\\
\hline
$\eta_c+\eta_c$        &(1.83 $\pm$ 0.10)$\times10^{-3}$
\\
$\eta_c+\eta_c(2S)$    &(1.52 $\pm$ 0.08)$\times10^{-3}$
\\
$\eta_c(2S)+\eta_c(2S)$&(0.31 $\pm$ 0.02)$\times10^{-3}$
\end{tabular}
\end{ruledtabular}
\end{table}

Our predictions for the double-charmonium cross sections for $S$-wave
states are given in Table~\ref{tab:sigmaSS}. The error bars are those
associated with the uncertainty in the pole mass $m_c$ only. The small
error bars for $\eta_c + \eta_c$ in Table~\ref{tab:sigmaSS} are a
consequence of the value of $m_c$ being fortuitously close to a zero in
the derivative of the cross section with respect to $m_c$. The cross
section for $\eta_c + \eta_c$ is about 3 orders of magnitude smaller
than that for $J/\psi + \eta_c$.  This suppression comes primarily from
the coupling-constant factor $\alpha^2/\alpha_s^2$. The cross section
for $J/\psi + J/\psi$ is larger than that for $J/\psi + \eta_c$. The
suppression factor of $\alpha^2/\alpha_s^2$ is more than compensated by
the kinematic enhancement factor that scales as $r^{-6}$. The cross
section for $J/\psi + J/\psi$ is dominated by the photon-fragmentation
diagrams in Figs.~\ref{fig1}(a) and \ref{fig1}(b). For $m_c=1.4$~GeV,
they contribute about 115\% of the cross section. In
Ref.~\cite{Braaten:2002fi}, it was pointed out that there may be large
perturbative-QCD and relativistic corrections to the production cross
section for $J/\psi + \eta_c$.  There may also be large perturbative-QCD
and relativistic corrections to the cross section for $J/\psi + J/\psi$.
These corrections would affect not only the absolute cross sections in
Table~\ref{tab:sigmaSS}, but also the ratios of cross sections. For the
$J/\psi+J/\psi$, $J/\psi+\psi(2S)$, and $\psi(2S)+\psi(2S)$ cross
sections, the dominant photon-fragmentation contributions receive
perturbative-QCD correction factors of about $0.39$ and relativistic
correction factors of about $0.78$, $0.62$, and $0.49$,
respectively~\cite{Bodwin:2002kk}.

\begin{figure}[ht]
\includegraphics[width=8cm]{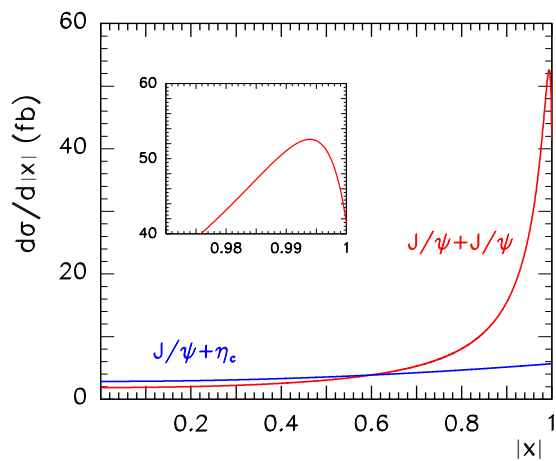}
\caption{\label{fig2}
Differential cross sections $d\sigma/d|x|$ for 
$e^+ e^-$ annihilation into $J/\psi+J/\psi$ and $J/\psi+\eta_c$
at $E_{\rm beam} = 5.3$ GeV.  
The areas under the curves 
are the integrated cross sections 6.65 fb  and 3.78 fb.
There are large errors associated with QCD and relativistic
corrections, as described in the text.
}
\end{figure}

The angular distributions $d \sigma/d|x|$ for $m_c = 1.4$ GeV are shown
in Fig.~\ref{fig2}. At $x=0$, the differential cross section for
$J/\psi+J/\psi$ (normalized as in Fig.~\ref{fig2}) is smaller
than that for $J/\psi+\eta_c$ by about a factor 0.66. However, the
differential cross section for $J/\psi+J/\psi$ is strongly peaked near
the beam direction at $x = 0.994$, where it is larger than that for
$J/\psi+\eta_c$ by about a factor 9.3. The reason for the sharp peak
is evident from the denominator in Eq.~(\ref{dRdx:psi+psi}).

The Belle Collaboration has recently measured the cross section for
$J/\psi + \eta_c$ by observing a peak in the momentum spectrum of the
$J/\psi$ that corresponds to the 2-body final state $J/\psi + \eta_c$
\cite{Abe:2002rb}. The measured cross section is
\begin{eqnarray}
\sigma[J/\psi+\eta_c] \times B[\ge 4] 
= \left( 33^{+7}_{-6} \pm 9 \right) \; {\rm fb},
\label{Belle}
\end{eqnarray}
where $B[\ge 4]$ is the branching fraction for the $\eta_c$ to decay
into at least 4 charged particles.  Since $B[\ge 4]<1$, the right side
of Eq.~(\ref{Belle}) is a lower bound on the cross section to produce
$J/\psi + \eta_c$. This lower bound is about an order of magnitude
larger than the predictions of NRQCD in the nonrelativistic limit
\cite{Braaten:2002fi,Liu:2002wq}. Large relativistic corrections may
account for part of the discrepancy \cite{Braaten:2002fi}. There may
also be large nonperturbative corrections associated with the
double-charmonium cross section being a significant fraction of the
total $c \bar c + c \bar c$ cross section \cite{Liu:2002wq}. However the
large discrepancy between the NRQCD predictions and the Belle
measurement is still disturbing. This discrepancy can be decreased by
taking into account the process $e^+ e^- \to J/\psi + J/\psi$. In the
Belle fit to the $J/\psi$ momentum distribution, the full width at half
maximum of the $\eta_c$ peak is about 0.11 GeV. Since the mass
difference between the $J/\psi$ and the $\eta_c$ is about 0.12 GeV,
there are probably $J/\psi + J/\psi$ events that contribute to the
$J/\psi + \eta_c$ signal observed by Belle.

The Belle Collaboration also saw evidence for $J/\psi + \chi_{c0}(1P)$
and $J/\psi + \eta_c(2S)$ events. The $\eta_c(2S)$ was recently
discovered by the Belle Collaboration at a mass $M_{\eta_c(2S)} = 3654
\pm 6 \pm 8$ MeV~\cite{Choi:2002na}. Since the mass difference between
the $\psi(2S)$ and the $\eta_c(2S)$ is only about 0.03 GeV, it is likely
that any signal for $J/\psi + \eta_c(2S)$ in the $J/\psi$ momentum
spectrum is contaminated by $J/\psi + \psi(2S)$ events. A 3-peak fit to
the data for the momentum spectrum of the $J/\psi$ gives approximately
67, 39, and 42 events with an accompanying $\eta_c$, $\chi_{c0}(1P)$, or
$\eta_c(2S)$, with an uncertainty of 12--15 events for each final state.
 The predictions in Ref.~\cite{Braaten:2002fi} for the relative cross
sections for production of a $J/\psi$ with an accompanying $\eta_c$,
$\chi_{c0}(1P)$, or $\eta_c(2S)$ are 1.00, 0.63, 0.42, respectively.
The observed proportion of events is 
compatible with the NRQCD predictions. 

There are also some unresolved puzzles in the inclusive $J/\psi$ cross
sections that have been measured by the Belle and \babar Collaborations
\cite{Abe:2001za,Aubert:2001pd}. There are significant discrepancies
between their measurements of both the production-angle distributions of
the $J/\psi$ and the $J/\psi$ polarization. There are also significant
discrepancies between the measurements and predictions based on NRQCD
\cite{Jpsi-X}. These discrepancies may be decreased by taking into
account photon-fragmentation contributions to the process $e^+ e^- \to
J/\psi + q \bar q$, which allow the order-$\alpha^4$ QED contribution to
compete with the order-$\alpha^2 \alpha_s^2$ QCD contribution. The two
experiments impose different cuts to decrease the background from the
two-photon processes $e^+ e^- \to e^+ e^- \gamma^* \gamma^*$ and from
initial-state-radiation processes such as $e^+ e^- \to \psi(2S) +
\gamma$ and $e^+ e^- \to J/\psi + \ell^+ \ell^-$. These cuts remove
substantial parts of the photon-fragmentation contribution to $J/\psi +
q \bar q$. The differences between the experiments could be due to
differences between the cuts. The discrepancies with the NRQCD
predictions could be due partly to the neglect of the order-$\alpha^4$
QED terms in the theoretical predictions.

Another puzzling result comes from the measurement by the Belle
Collaboration of the fraction of the inclusive $J/\psi$ cross section
that is attributable to the final state $J/\psi+ c \bar c$
\cite{Abe:2002rb}. That fraction is much larger than predictions based
on NRQCD. The discrepancies between the predictions and the measurements
would be decreased by taking into account the contributions to the
processes $e^+ e^- \to J/\psi + q \bar q$ and $e^+ e^- \to J/\psi + c
\bar c$ from photon fragmentation into $J/\psi$. The experimental cuts
allow the photon-fragmentation contribution to $J/\psi + c \bar c$ to
survive, but they remove substantial parts of the photon-fragmentation
contribution to $J/\psi + q \bar q$. Thus, the net effect of the
photon-fragmentation processes is to increase the ratio of $J/\psi+ c
\bar c$ events to $J/\psi+ X$ events. 

In summary, we have calculated the cross section for $e^+ e^-$
annihilation into $J/\psi + J/\psi$.  The calculated cross section is
larger than that for $e^+ e^-\to J/\psi + \eta_c$ by about a factor of
$1.8$ because it receives contributions from the process in which two
virtual photons fragment into two $c\bar c$ pairs. The inclusion of this
process in the analysis may decrease the large discrepancy between the
Belle measurement of the production cross section for $J/\psi + \eta_c$
and the predictions based on NRQCD.

One of us (E.B.) would like to thank B.~Yabsley for valuable discussions.
Research in the HEP Division at Argonne National Laboratory is supported
by the U.~S.~Department of Energy under Contract W-31-109-ENG-38.
Fermilab is operated by Universities Research Association Inc.~under
Contract DE-AC02-76CH03000 with the Department of Energy. The research
of E.B.~is also supported in part by the Department of Energy under
grant DE-FG02-91-ER4069.


\end{document}